\pdfoutput=1
\documentclass[12pt,a4paper]{article}

\usepackage{ifthen} 
\newboolean{pdflatex}
\setboolean{pdflatex}{true} 

\newboolean{articletitles}
\setboolean{articletitles}{true} 

\newboolean{uprightparticles}
\setboolean{uprightparticles}{false} 

\usepackage{pdfpages} 

\def\paperauthors{LHCb collaboration} 
\def\paperasciititle{Computing and software for LHCb Upgrade II} 
\def\papertitle{Computing and software for LHCb Upgrade~II} 
\def\paperkeywords{ {LHCb}} 
\def\papercopyright{\the\year\ CERN for the benefit of the LHCb collaboration} 
\def\paperlicence{CC BY 4.0 licence}
\def\paperlicenceurl{https://creativecommons.org/licenses/by/4.0/}

\usepackage[top=1in, bottom=1.25in, left=1in, right=1in]{geometry}

\usepackage{microtype}
\usepackage{lineno}  
\usepackage{xspace} 
\usepackage{caption} 

\usepackage{graphicx}  
\usepackage{color}
\usepackage{colortbl}
\graphicspath{{./figs/}} 

\usepackage{amsmath} 
\usepackage{amssymb}
\usepackage{amsfonts}
\usepackage{upgreek} 

\usepackage{multirow}

\newcommand*\patchAmsMathEnvironmentForLineno[1]{%
\expandafter\let\csname old#1\expandafter\endcsname\csname #1\endcsname
\expandafter\let\csname oldend#1\expandafter\endcsname\csname
end#1\endcsname
 \renewenvironment{#1}%
   {\linenomath\csname old#1\endcsname}%
   {\csname oldend#1\endcsname\endlinenomath}%
}
\newcommand*\patchBothAmsMathEnvironmentsForLineno[1]{%
  \patchAmsMathEnvironmentForLineno{#1}%
  \patchAmsMathEnvironmentForLineno{#1*}%
}
\AtBeginDocument{%
\patchBothAmsMathEnvironmentsForLineno{equation}%
\patchBothAmsMathEnvironmentsForLineno{align}%
\patchBothAmsMathEnvironmentsForLineno{flalign}%
\patchBothAmsMathEnvironmentsForLineno{alignat}%
\patchBothAmsMathEnvironmentsForLineno{gather}%
\patchBothAmsMathEnvironmentsForLineno{multline}%
\patchBothAmsMathEnvironmentsForLineno{eqnarray}%
}

\usepackage[pdftex,
            pdfauthor={\paperauthors},
            pdftitle={\paperasciititle},
            pdfkeywords={\paperkeywords},
]{hyperref}

\usepackage{hyperxmp}

\hypersetup{pdfcopyright={Copyright (C) \papercopyright}}
\hypersetup{pdflicenseurl={\paperlicenceurl}}

\usepackage[bottom,flushmargin,hang,multiple]{footmisc}

\usepackage[all]{hypcap} 

\usepackage{xspace} 
\usepackage{upgreek}

\def\lhcb   {\mbox{LHCb}\xspace}

\def\MagUp {\mbox{\em Mag\kern -0.05em Up}\xspace}

\ifthenelse{\boolean{uprightparticles}}%
{

 \def\PDelta      {\ensuremath{\Delta}\xspace}                 
 \def\PXi         {\ensuremath{\Xi}\xspace}                 
 \def\PLambda     {\ensuremath{\Lambda}\xspace}                 
 \def\PSigma      {\ensuremath{\Sigma}\xspace}                 
 \def\POmega      {\ensuremath{\Omega}\xspace}                 
 \def\PUpsilon    {\ensuremath{\Upsilon}\xspace}
 \let\oldPi\Pi
 \def\PPi         {\ensuremath{\oldPi}\xspace}

 \def\PB      {\ensuremath{\mathrm{B}}\xspace}                 
                  
 \def\PD      {\ensuremath{\mathrm{D}}\xspace}

 \def\PK      {\ensuremath{\mathrm{K}}\xspace}

 \def\Pi      {\ensuremath{\mathrm{i}}\xspace}

 \def\Ps      {\ensuremath{\mathrm{s}}\xspace}

 \def\thebaroffset{0.0em}
}
{

 \mathchardef\PDelta="7101
 \mathchardef\PXi="7104
 \mathchardef\PLambda="7103
 \mathchardef\PSigma="7106
 \mathchardef\POmega="710A
 \mathchardef\PUpsilon="7107
 \mathchardef\PPi="7105
                  
 \def\PB      {\ensuremath{B}\xspace}                 
                  
 \def\PD      {\ensuremath{D}\xspace}

 \def\PK      {\ensuremath{K}\xspace}

 \def\Pi      {\ensuremath{i}\xspace}

 \def\Ps      {\ensuremath{s}\xspace}

 \def\thebaroffset{0.18em}
}
\newcommand{\offsetoverline}[2][\thebaroffset]{\kern #1\overline{\kern -#1 #2}}%

\makeatletter
\ifcase \@ptsize \relax
  \newcommand{\miniscule}{\@setfontsize\miniscule{4}{5}}
\or
  \newcommand{\miniscule}{\@setfontsize\miniscule{5}{6}}
\or
  \newcommand{\miniscule}{\@setfontsize\miniscule{5}{6}}
\fi
\makeatother

\DeclareRobustCommand{\optbar}[1]{\shortstack{{\miniscule (\rule[.5ex]{1.25em}{.18mm})}
  \\ [-.7ex] $#1$}}

\def\squark    {{\ensuremath{\Ps}}\xspace}

\def\KorKbar {\kern \thebaroffset\optbar{\kern -\thebaroffset \PK}{}\xspace}

\def\D       {{\ensuremath{\PD}}\xspace}

\def\DorDbar {\kern \thebaroffset\optbar{\kern -\thebaroffset \PD}\xspace}

\def\Dp      {{\ensuremath{\D^+}}\xspace}
\def\Dm      {{\ensuremath{\D^-}}\xspace}

\def\DpDm    {\ensuremath{\Dp {\kern -0.16em \Dm}}\xspace}

\def\B       {{\ensuremath{\PB}}\xspace}

\def\BorBbar {\kern \thebaroffset\optbar{\kern -\thebaroffset \PB}\xspace}

\def\Bd      {{\ensuremath{\B^0}}\xspace}

\def\BdorBdbar {\kern \thebaroffset\optbar{\kern -\thebaroffset \Bd}\xspace}

\def\Bs      {{\ensuremath{\B^0_\squark}}\xspace}

\def\BsorBsbar {\kern \thebaroffset\optbar{\kern -\thebaroffset \Bs}\xspace}

\def\Y#1S{\ensuremath{\PUpsilon{(#1S)}}\xspace}

\def\LorLbar     {\kern \thebaroffset\optbar{\kern -\thebaroffset \PLambda}\xspace}

\def\AT#1     {\ensuremath{A_{\mathrm{T}}^{#1}}\xspace}           

\def\C#1      {\ensuremath{\mathcal{C}_{#1}}\xspace}                       
\def\Cp#1     {\ensuremath{\mathcal{C}_{#1}^{'}}\xspace}                    
\def\Ceff#1   {\ensuremath{\mathcal{C}_{#1}^{\mathrm{(eff)}}}\xspace}        
\def\Cpeff#1  {\ensuremath{\mathcal{C}_{#1}^{'\mathrm{(eff)}}}\xspace}       
\def\Ope#1    {\ensuremath{\mathcal{O}_{#1}}\xspace}                       
\def\Opep#1   {\ensuremath{\mathcal{O}_{#1}^{'}}\xspace}                    

\newcommand{\aunit}[1]{\ensuremath{\text{\,#1}}}       

\newcommand{\tev}{\aunit{Te\kern -0.1em V}\xspace}
\newcommand{\gev}{\aunit{Ge\kern -0.1em V}\xspace}
\newcommand{\mev}{\aunit{Me\kern -0.1em V}\xspace}
\newcommand{\kev}{\aunit{ke\kern -0.1em V}\xspace}
\newcommand{\ev}{\aunit{e\kern -0.1em V}\xspace}
 
\newcommand{\mevc}{\ensuremath{\aunit{Me\kern -0.1em V\!/}c}\xspace}
\newcommand{\gevc}{\ensuremath{\aunit{Ge\kern -0.1em V\!/}c}\xspace}
\newcommand{\mevcc}{\ensuremath{\aunit{Me\kern -0.1em V\!/}c^2}\xspace}
\newcommand{\gevcc}{\ensuremath{\aunit{Ge\kern -0.1em V\!/}c^2}\xspace}

\def\cm   {\aunit{cm}\xspace}

\def\sec  {\ensuremath{\aunit{s}}\xspace}

\def\gsim{{~\raise.15em\hbox{$>$}\kern-.85em
          \lower.35em\hbox{$\sim$}~}\xspace}
\def\lsim{{~\raise.15em\hbox{$<$}\kern-.85em
          \lower.35em\hbox{$\sim$}~}\xspace}

\def\geant      {\mbox{\textsc{Geant4}}\xspace}

\def\pythia     {\mbox{\textsc{Pythia}}\xspace}

\def\tell1  {TELL1\xspace}
\def\ukl1   {UKL1\xspace}

\newcommand{\lhcborcid}[1]{\href{https://orcid.org/#1}{\hspace*{0.1em}\raisebox{-0.45ex}{\includegraphics[width=1em]{figs/orcidIcon.pdf}}}}

\usepackage{cite} 
\usepackage{mciteplus}

\usepackage{comment}
\usepackage{tikz}
\usetikzlibrary{shapes}
\usepackage{pgfgantt}
\usepackage{pdflscape}

\begin{document}


\renewcommand{\thefootnote}{\fnsymbol{footnote}}
\setcounter{footnote}{1}

\begin{titlepage}
\pagenumbering{roman}

\vspace*{-1.5cm}
\centerline{\large EUROPEAN ORGANIZATION FOR NUCLEAR RESEARCH (CERN)}
\vspace*{1.5cm}
\noindent
\begin{tabular*}{\linewidth}{lc@{\extracolsep{\fill}}r@{\extracolsep{0pt}}}
\ifthenelse{\boolean{pdflatex}}
{\vspace*{-1.5cm}\mbox{\!\!\!\includegraphics[width=.14\textwidth]{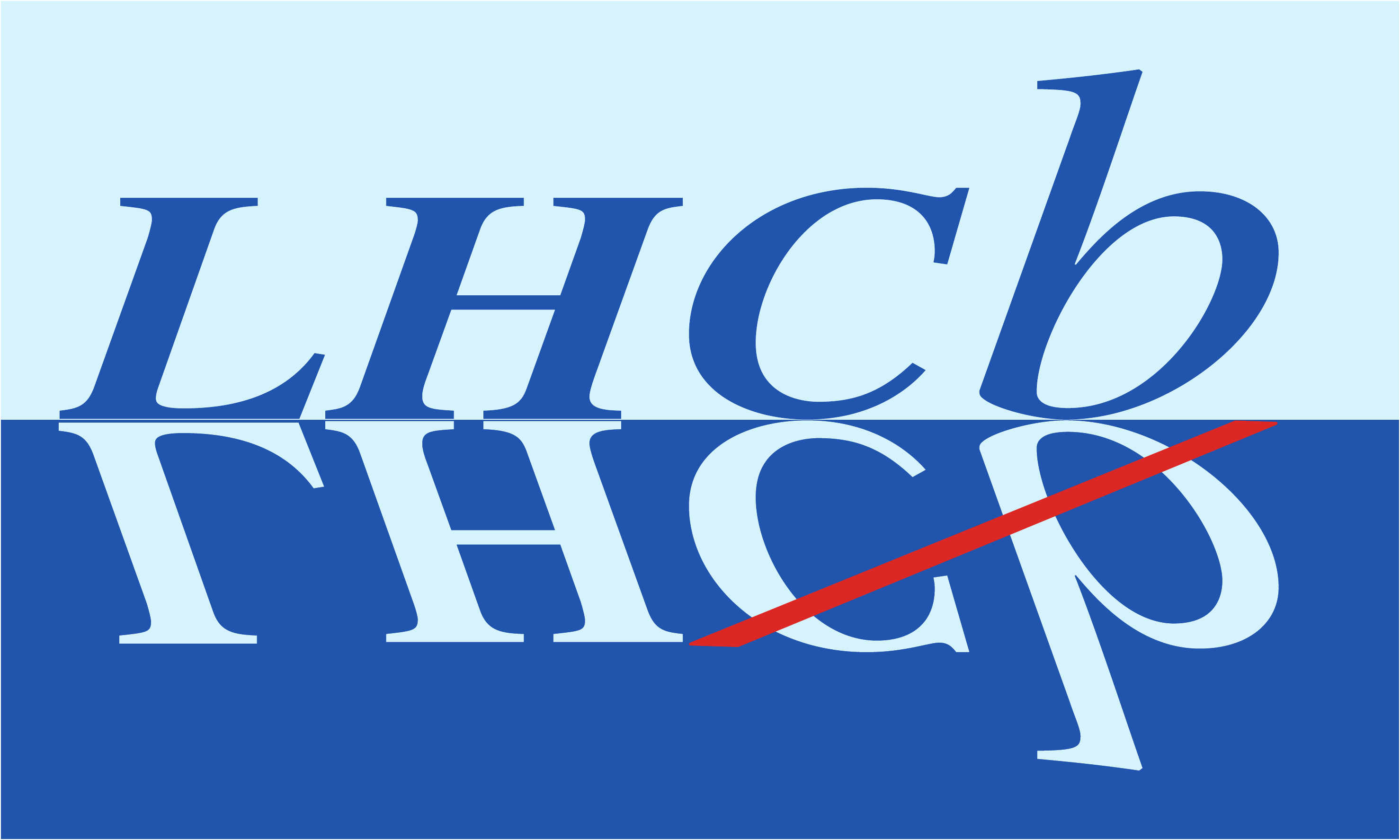}} & &}%
{\vspace*{-1.2cm}\mbox{\!\!\!\includegraphics[width=.12\textwidth]{figs/lhcb-logo.eps}} & &}%
\\
 & & LHCb-PUB-2025-004 \\  
 & & \today \\ 
 & & \\
\end{tabular*}

\vspace*{2.0cm}

{\normalfont\bfseries\boldmath\huge 
\begin{center}
    \papertitle \\
    \vspace*{0.5cm}
  {\normalsize Input to the European Particle Physics Strategy Update 2024--26}
\end{center}
}

\vspace*{1.0cm}

\begin{center}
\paperauthors\footnote{
    Contact authors: 
    Vincenzo Vagnoni (\href{mailto:vincenzo.vagnoni@cern.ch}{vincenzo.vagnoni@cern.ch}),
    Patrick Robbe (\href{mailto:patrick.robbe@cern.ch}{patrick.robbe@cern.ch}),
    Ben Couturier (\href{mailto:ben.couturier@cern.ch}{ben.couturier@cern.ch}),
    Carla Marin Benito (\href{mailto:carla.marin.benito@cern.ch}{carla.marin.benito@cern.ch}).
}
\end{center}

\vspace{\fill}

\begin{abstract}
    \noindent
    A second major upgrade of the LHCb experiment is necessary to allow full exploitation of the High Luminosity LHC for flavour physics. The new experiment will operate in Run 5 of the LHC at a luminosity up to $1.5\times 10^{34}\cm^{-2}\sec^{-1}$. The experiment will therefore experience extremely high particle fluences and data rates, posing a high challenge not only for the detector but also for the software and computing resources needed to readout, reconstruct, select and analyse the data. This document presents these challenges and the ongoing and future R\&D programme necessary to address them. This programme will benefit not only the LHCb Upgrade II experiment, but the whole particle physics community as similar challenges will be faced by the next generation of experiments.

\end{abstract}

\vspace{\fill}

{\footnotesize 
\centerline{\copyright~\papercopyright. \href{\paperlicenceurl}{\paperlicence}.}}
\vspace*{2mm}

\end{titlepage}


\newpage
\setcounter{page}{2}
\mbox{~}

\renewcommand{\thefootnote}{\arabic{footnote}}
\setcounter{footnote}{0}



\cleardoublepage


\pagestyle{plain} 
\setcounter{page}{1}
\pagenumbering{arabic}



\section{Introduction}
The LHCb collaboration successfully demonstrated the feasibility and advantages of a heavy-flavour experiment at an energy frontier hadron collider during Runs~1 and~2 of the LHC. This motivated an effort to upgrade the experiment’s capabilities, to allow full exploitation of the unprecedented rate of heavy quarks that LHC collisions produce at full luminosity. While the initial experiment ran at an instantaneous luminosity of $2\times 10^{32}\cm^{-2}\sec^{-1}$, the Upgrade~I of the detector~\cite{LHCb-DP-2022-002} was designed for and has recently been proven to work at $2\times 10^{33}\cm^{-2}\sec^{-1}$.

The LHCb collaboration is now planning for an increase of the luminosity by another order of magnitude for Run 5 of the LHC, aiming at more than $10^{34}\cm^{-2}\sec^{-1}$, with its Upgrade~II project. 
Precision flavour physics in high pile-up hadron collisions is a special challenge for this project that requires not only advancement in detector technology -- including precise measurement of the timing of particles~\cite{LHCb-PUB-2025-002} -- but also significant developments in software and computing, ranging from extremely good pattern recognition in a crowded environment to highly sophisticated and custom-designed data processing and huge amounts of storage.

At the instantaneous luminosity of the Upgrade~II LHCb experiment, each bunch crossing will produce on average two charm hadrons in the LHCb acceptance and more than one in ten will produce a beauty hadron~\cite{LHCB-TDR-026}. While this allows a unique flavour physics programme, it requires reconstructing every event, storing a very large fraction of them, analysing unprecedented amounts of data offline, and simulating an enormous amount of complex events.

\begin{figure}
    \centering
    \includegraphics[width=0.8\linewidth]{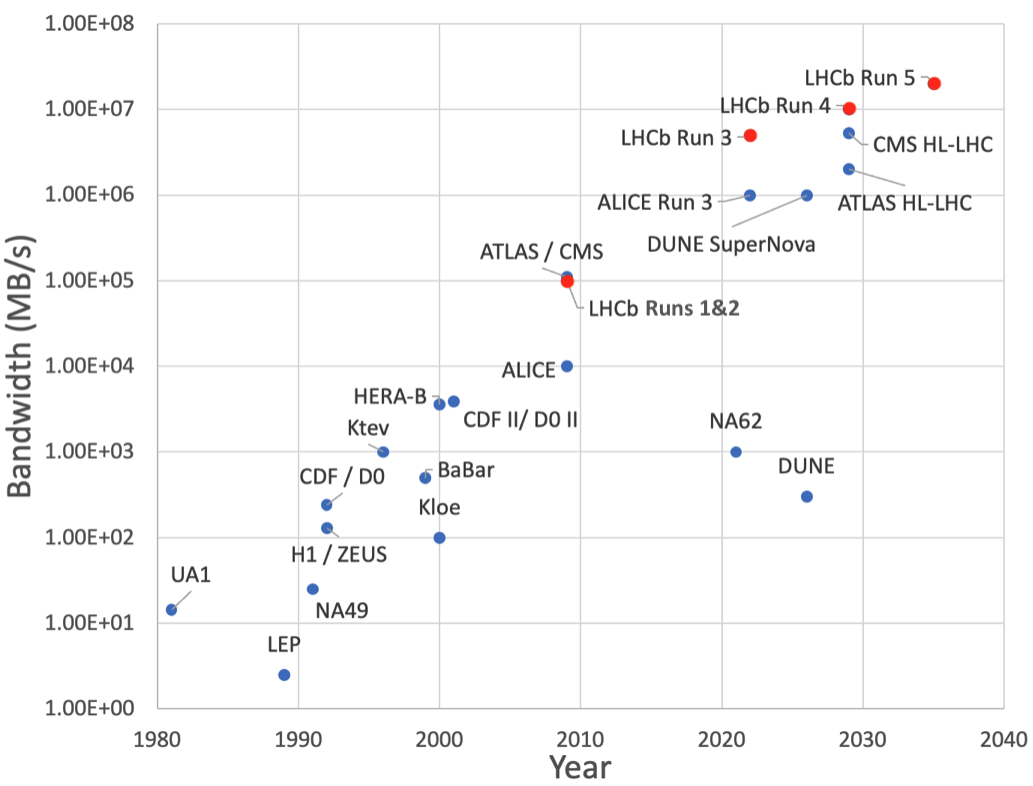}
    \caption{Evolution of experiment bandwidths in High Energy Physics as a function of time [graph
courtesy of A. Cerri, U. of Sussex]. The various LHCb experiment phases are highlighted in red.}
    \label{fig:BW-Cerri}
\end{figure}

The Upgrade~I experiment has set the ground to overcome such a challenge, proving the viability of a trigger-less readout, a fully software-based trigger and a selective persistence model to minimise the amount of information stored per event. However, further developments will be required to enable operation at a much higher luminosity. The total data bandwidth expected in Upgrade~II exceeds $10^7~\text{MB/s}$~\cite{LHCB-TDR-026}, surpassing significantly what is planned for the High-Luminosity LHC upgrades of the ATLAS~\cite{CERN-LHCC-2020-015} and CMS~\cite{CERN-LHCC-2015-010} experiments, which will continue using hardware triggers to reduce the amount of data to be readout from the detector and processed in software. Figure~\ref{fig:BW-Cerri} shows the evolution of the data bandwidth as a function of time for past and planned experiments. LHCb Run~5 will be the biggest data processing challenge ever faced in High Energy Physics~(HEP) and will consequently act as a pathfinder for any future collider experiments.

The present document provides an executive summary of the main directions of software and computing development ongoing within the context of LHCb Upgrade~II. It is organized
in three broad areas: data acquisition, software trigger and reconstruction in Section~\ref{sec:daq_trigger}, simulation in Section~\ref{sec:simulation} and offline data processing and analysis in Section~\ref{sec:offline}. Closing remarks are presented in Section~\ref{sec:closing}. Full details about the Upgrade~II project can be found in the Framework TDR~\cite{LHCB-TDR-023} and the Scoping Document~\cite{LHCB-TDR-026}.

\section{Data acquisition, software trigger and reconstruction}
\label{sec:daq_trigger}

In order to fully exploit the physics potential of the experiment, the LHCb data acquisition (DAQ) system has been completely redesigned for Upgrade~I. This redesign follows a trigger-less design that performs a full readout of the detector at the LHC collision rate. This redesign has been made possible by leveraging significant advancements in commercial off-the-shelf (COTS) network technologies, as well as in the computing power and memory bandwidth of modern server platforms. These advancements have been further supported by clever and efficient software design. This innovative approach has enabled LHCb to achieve an unprecedented volume of data processing in the HEP field. For the Upgrade~II of the LHCb experiment, the strategy will be to continue along the trajectory established during the first upgrade, placing even greater demands upon the DAQ system. Given the time frame of the upgrade, it is possible to leverage the progress in COTS technologies and the considerable experience in effective and efficient software design gained during the first upgrade.

The present DAQ system is a heterogeneous computing platform comprising field-programmable gate array (FPGA)-based DAQ cards, high-speed dedicated network cards, and graphics processing units (GPUs) for data processing. This heterogeneous architecture makes it imperative to design and implement software in a way that leverages the capabilities of the complete system. The utilization of FPGA cards in this system is particularly advantageous due to their efficacy in low-level bit manipulations, which, on conventional CPU/GPU architectures, are notoriously computationally expensive. Modern high-throughput network cards are capable of handling most memory operations necessary for data transfer, while GPUs are optimized for solving large-scale parallel problems, such as particle reconstruction. The optimal distribution of workload across these components necessitates a profound comprehension of their underlying hardware architectures. During the design, R\&D, and commissioning phases of the Run 3 experiment, the LHCb collaboration gained significant experience in all the details of such an heterogeneous system by putting together a multidisciplinary team of physicists, engineers, and computer scientists. The maintenance and expansion of such a team is crucial for the success of the Upgrade~II of the experiment and for the progress of the whole HEP environment.

During the design phase, it is preferable to implement a global cost modelling and optimization approach rather than to over-optimize individual components in isolation, as this can compromise the system's overall efficiency. Evaluating the performance of such a complex system often requires access to a substantial fraction of the final hardware setup. However, achieving this during the R\&D phase is challenging due to the fast technology evolution, and limited budgets available for infrastructure and testing. The lack of access to a suitable test infrastructure significantly impacts the evaluation of network scalability. This poses significant challenges in the design of a DAQ system. Low-level network simulations can have a considerable impact on the decision-making process and consequently influence the final decision regarding the DAQ architecture. However, developing an accurate low-level model of a large network infrastructure is a highly challenging task, requiring substantial computing resources to achieve meaningful results. In this regard, the LHCb collaboration is well-positioned to lead the effort to develop open-source low-level simulation models for different network technologies and to foster collaboration across the HEP community and beyond. Subsequent phases will involve leveraging external resources, such as high-performance computing (HPC) sites, to further refine our models and accelerate progress. During the first upgrade of the experiment, for instance, access to HPC installations equipped with suitable network technologies was critical for large-scale performance testing and evaluation. Ensuring the availability of such resources in the future will be essential. This can be achieved through collaborations with European HPC sites or as a shared facility for the HEP community.

The primary risk of relying on COTS hardware lies in the potential unavailability or obsolescence of specific components. To mitigate this, we designed a software architecture that supports rapid adaptation to emerging hardware technologies. Maintaining multiple options for the network can accelerate these transitions and reduce risks associated with vendor lock-in or technological obsolescence. Initiatives such as the Ultra Ethernet consortium~\cite{10614558}, which aims to design next-generation interconnection technologies, warrant close attention in this regard.
Programming languages used in the project should also be periodically reassessed. Incorporating modern languages like Julia or Rust can significantly reduce the software development and maintenance effort, due to their built-in functionalities. This can streamline certain aspects of the software, such as data handling over the network, storage system management, and implementation of simple analysis tasks such as online data monitoring. Dedicated training is crucial to maintain a pool of experts with the necessary skills and knowledge to exploit such modern languages.

The core goal of LHCb's online data processing is to select the broad range of physics processes that LHCb wishes to study, from very low momentum $K_{\rm S}^0$ and $\mathit{\Lambda}$ hadrons up to electroweak bosons. At the same time, it must reduce the enormous background from nonsignal processes produced in hadronic collisions.
To achieve this, LHCb has led the R\&D of trigger system design since the start of the LHC, gradually removing stages towards a general purpose software trigger operating at the collision frequency.
From the original concept, consisting of a true hardware trigger (also referred to as Level 0) with reduced readout rate, the system has evolved into a multi-stage fully-software based trigger (also referred to as High Level Trigger or HLT).

In Run 3, LHCb has deployed an innovative trigger system that allows us to readout the detector, reconstruct tracks and filter events in software at the LHC bunch crossing rate, a first for a collider physics experiment~\cite{Aaij:2019zbu}. 
To accomplish this, LHCb has pioneered the usage of modern technologies in ways that are new to the field.
The first software trigger stage (HLT1) has been rewritten from the ground up to run on GPUs as these were proven to reduce the total cost for the same performance and offer reduced energy consumption~\cite{LHCb-DP-2021-003}. In HLT1, a complete tracking sequence and a subset of the full particle identification is performed exploiting the highly parallel capabilities of GPUs and providing functionality beyond the original TDR design~\cite{LHCb-TDR-016}. 
Moreover, the HLT1 framework allows scalable pipelines mixing classical and Machine-Learning (ML)/Artificial-Intelligence (AI) algorithms to be deployed on a range of parallel architectures~\cite{Aaij:2019zbu}.

Additional requirements on the LHCb trigger system are imposed by the demands of high precision measurements. To start with, it is important to minimize differences between online (HLT) and offline selections as these are inherent sources of bias in precision measurements. Already in Run~1, LHCb deployed quantised ML methods in the main trigger selections in order to make them as robust as possible against changing detector conditions~\cite{BBDT}.
In Run~3, this demand is satisfied by performing the full, offline-quality event reconstruction using all subdetectors at the second software trigger stage (HLT2), which uses the x86 architecture to process events.
This allows events to be selected based on physics quantities rather than approximations to these. A disk buffer between the first and second trigger stages enables a real-time alignment and calibration of the detectors during data taking, the outcome of which is used when events are reconstructed at HLT2, providing the best physics quality prior to HLT2 selections.

Moreover, detailed information must be stored for specific decays, typically rare signals, while very large rates are needed for the control and calibration modes. The ability to prescale or write partial event information for Cabibbo-favoured charm decays while saving additional information for suppressed modes is one such example, where the sensitivity to physics observables is driven by the rarer process~\cite{LHCb-DP-2019-001}. However, the output data volume is limited to the available storage resources. 
LHCb's innovative solution consists of providing full flexibility on the amount of information stored offline for each trigger selection. 
Performing the full, offline-quality reconstruction in HLT2 enables complex selections exploiting not only kinematic and topological criteria but also particle identification information. This allows to  distinguish and stream topologically similar final states in different (reduced) data formats to offline.
This system was introduced already in Run~2  with a successful outcome~\cite{LHCb-PAPER-2019-006}. In Run~3, HLT2 provides the final full-quality reconstruction, completely mitigating the need to rerun the reconstruction offline, which additionally saves offline computing resources.

The LHCb Upgrade~II programme will increase the instantaneous luminosity by an order of magnitude, increasing the signal rates in the same proportion. A full software trigger, enabling selection and classification of different signal topologies and flexible storage of event information at the trigger level will continue to be mandatory in order to store the plethora of signal events that will be produced. 
However, the increase in luminosity comes at the cost of enormous combinatorics when reconstructing tracks, vertices and particle identification information prior to making selections. The use of timing information in the reconstruction algorithms will be necessary to separate information from different pile-up collisions. This will require the modification of the whole reconstruction sequence to include timing information from several detectors.
The usage of ML/AI strategies to combine timing and position information is a promising approach that builds on the track record of LHCb exploiting such algorithms. As an example, Neural Networks (NNs), including like-for-like comparisons of shallow and deep NNs, have been used in the tracking algorithms from Run~2 onwards.

Additionally, the computational challenge of reading out the full detector, reconstructing tracks and filtering events at the bunch crossing rate will be unprecedented in LHCb Upgrade~II, as shown in Fig.~\ref{fig:BW-Cerri}.  
The baseline solution relies on moving the entire reconstruction to GPUs. However, other technologies may be available at competitive cost efficiency in the coming years, and LHCb will develop its data processing framework such that the real-time reconstruction and selections can be easily adapted to the best suitable technology. The expertise developing a GPU-based framework~\cite{Aaij:2019zbu} for the Run~3 HLT1 together with the experience developing and operating a CPU-based framework~\cite{Gaudi} since Run~1, are already being exploited to design a flexible system capable of supporting different architectures.

The implementation of the online alignment and calibration of the detector will also need to be revisited in Upgrade~II: the time available to carry out these processes will be constrained by the data volume of the disk buffer in between HLT1 and HLT2. Moreover, a more complicated alignment process incorporating timing-based reconstruction will be needed. Exploiting the parallel capabilities of GPUs for the alignment processes is a promising direction, based on the expertise of LHCb using this technology.

Furthermore, the lack of a low-level hardware trigger imposes additional demands on the online monitoring infrastructure. This system, essential for assessing detector conditions, is a major component already in Run~3. The computing power available for online monitoring has grown considerably since the beginning of LHC data-taking. Therefore an update of the software infrastructure is envisaged for Upgrade~II, to ensure an optimal use of the available resources. Task scheduling on suitable hardware platforms presents challenges akin to generic scheduling on general-purpose computing infrastructure. Leveraging and adapting industry-standard solutions rather than developing custom implementations can streamline deployment and optimize resource utilization. A similar approach can be applied to event transfer into the infrastructure used for the monitoring, further enhancing its efficiency and effectiveness.

Looking further ahead, future hadron collider designs will face similar issues to LHCb Upgrade~II at higher pileup. The evolution of LHCb's trigger from Run~1 to Upgrade~II serves as important R\&D for what a future collider detector trigger design will look like\cite{LHCb-DP-2019-002}. The challenges ahead are superlative but LHCb is placed in a unique position to address them, provided enough person power is available and the expertise developed over the first upgrade can be retained for future developments.



\section{Simulation}
\label{sec:simulation}

Simulated data are crucial for many aspects of the LHCb experiment, from supporting physics analysis of all available datasets to providing samples to design and study future upgrade projects. Production of simulated samples dominates the total offline CPU power required by the experiment, typically using around 90\% of available resources. The available CPU comes from the Worldwide LHC Computing Grid (WLCG), the LHCb online farm (outside of data taking periods) as well as some opportunistic resources. The contribution from the online farm should not be underestimated, it provides up to 50\% of available resources when the farm is not used to run the software trigger.

Looking ahead, it will be challenging to provide adequate simulated samples without using significantly more computing resources. This is because the upgraded LHCb detector has already collected more data during 2024 than was available from the whole of Run~1 and Run~2 combined. To keep systematic uncertainties from the limited size of simulated samples under control, simulation productions must keep up with the order of magnitude jumps in data taking for Run~3 and~4 and another subsequent step for Upgrade~II in Run~5. Larger data samples also imply that systematic uncertainties become relatively more important, meaning that the quality of the simulation should not be compromised, and if possible it should become more accurate. It is clear that the only realistic way forwards is to continue to significantly speed up the standard detailed simulation and use fast simulation options wherever possible.

Standard simulations at LHCb utilise existing generator 
 packages such as \pythia~\cite{Sjostrand:2007gs} 
 The detailed detector simulation is performed using the \geant\ toolkit~\cite{Allison:2006ve,*Agostinelli:2002hh}; this step typically requires the majority of the CPU time when producing simulated samples. For the Run 3 LHCb detector, approximately 50\% of the time spent in the \geant step is associated to the calorimeter simulation (dominated by the ECAL) with a further 25\% taken by the RICH detectors. This leads to the idea of fast simulations to reduce the amount of time it takes to simulate an event. 

Fast simulation techniques are already heavily used in LHCb. Since the introduction of ReDecay~\cite{Muller:2018vny} in 2019, around 75\% of simulated events come from fast simulation techniques, significantly above the current computing 
model expectations of 67\%~\cite{LHCb-TDR-018}. Tracker-only samples~\cite{Whitehead:2019bth}, where the calorimeters and RICH physics processes are turned off, have also been heavily used by analyses requiring very large samples of events.

 One important avenue for future progress will be to implement (parts of) the simulation workflow on GPUs, in particular to speed up processes such as optical photons and electromagnetic showers. The possibility of utilising the LHCb trigger GPU farm for simulation productions makes this avenue especially interesting. On the calorimetry side, work is ongoing to integrate both the AdePT prototype~\cite{Amadio:2022mqi} and the Celeritas code~\cite{Tognini:2022nmd} within the Gaussino framework~\cite{Mazurek_Corti_Müller_2021}, providing a fruitful synergy between external and LHCb R\&D projects. There are additional, more standalone, efforts to exploit the OPTICKS project~\cite{Blyth:2019yrd} and Mitsuba library~\cite{Jakob2020DrJit} to speed up optical photon simulation.
 
 The significant effort and developments required to ensure that simulation productions can be split between CPUs and GPUs simultaneously should not be underestimated. 
 Other architectures, such as ARM, have also been utilised for test productions. Samples were successfully produced and small differences when compared to CPU samples have now been resolved. This will enable us to take advantage of opportunistic ARM resources (or those pledged by WLCG sites) in the future. Supporting various architectures as soon as possible is critical to the adaptation of new technologies as they arise in the future.

Beyond fast simulation, the LHCb flash simulation is provided by the Lamarr package~\cite{LHCbSimulationProject:2023saz}, which aims to produce events around one thousand times faster than standard simulation. The idea is to replace the Geant4 simulation and event reconstruction phases with parameterisations of these, to be able to transform generator-level quantities directly into analysis-level reconstructed objects. It consists of a pipeline of modular parameterisations, trained on simulated samples and, where possible, real data, with separate branches for neutral and charged particles. The Lamarr pipeline, to produce Run~2 simulated samples, has recently been validated on three cloud computing sites across Italy. 

Further work is ongoing to provide fast simulation options for the calorimeters. The two approaches under investigation are to produce the calorimeter energy deposits either by using a point library~\cite{Rama:2019smm} or generative models, such as GANs~\cite{Chekalina:2018hxi}. First samples exploiting both of these techniques are expected to be produced in 2025. Both methods perform at least an order of magnitude faster than the standard simulation, so will provide excellent alternatives in the long term for Upgrade~II.

Additional challenges in the Upgrade~II era will include trying to speed up the event generators themselves, which may become the limiting factor given the efforts to speed up the detector simulation described above. While this is a community-wide effort driven by the generator developers, it is important that experiments are part of a testing and feedback loop. Generation and simulation of pile up, additional collisions in the considered bunch crossing, pose a significant challenge when scaling from five interactions per bunch crossing in Run~3 to around 40 in Run~5. Currently both hard and soft interactions are generated and mixed in the correct ratios, if this approach cannot scale, alternatives such as overlaying a random simulated or data background event can be considered. ReDecay may also be further exploited in this area.

\section{Offline data processing and analysis}
\label{sec:offline}

Offline processing of the data selected by the online trigger system is needed to further reduce the data volumes to a manageable level before analysis pipelines can be run.
In Run~3, offline processing begins with ``Sprucing", in which fully reconstructed events from HLT2 are analysed to reconstruct candidates for decay chains that were not considered in the real-time processing. At the same time, events are slimmed (removing information that is not required for analysis) such that they can be saved on disk and made available for further analysis steps.
Sprucing can additionally be used to compute derived physics quantities using algorithms too expensive to run in the real-time HLT.
The output data are split into streams of manageable size grouped by physics output, and saved on disk resources. This process can be rerun as physics selections change or improve, since the inputs are fully preserved on tape storage.

\lhcb analysts use WLCG resources through the Analysis Productions framework~\cite{Skidmore:2827261} to read the Sprucing output available on disk storage and create tuples that are used for their physics analyses. The analysts specify their decays of interest and which physical observables they would like regarding their decays, while the framework manages the submission of jobs and bookkeeping of input and output data. Additionally, some further reconstruction and combination steps can be performed if needed.

The primary challenges of offline data processing involve ensuring that there is enough tape storage for the expected HLT2 bandwidth over the data-taking period, and adequate disk storage availability for the Sprucing output, such that analysts are able to access the data on-demand. The output rates expected from Online processing and the Sprucing would reach $O(100)\,{\rm GB}/{\rm s}$ extrapolating the Run~3 model to the Upgrade~II luminosity and pile-up. New approaches will be mandatory to reduce the necessary resources, and the bandwidth should be carefully measured, and optimised where possible --- {\it e.g.} with more automated and robust monitoring.
An alternative strategy could be deferring the Sprucing step by ``parking" the input data on tape storage, a cheaper resource than disks, and running the Sprucing at a later time. Another option could be running  the Sprucing concurrently but storing the output directly to tape. However, these alternatives are inconvenient for the analysts, as data stored on tape is not available ``on-demand" to be processed by computing jobs at user level. To accommodate analysis preparation, a partial staging (copying data from tape to disk) could be carried out such that analysts are able to prepare their pipelines on a subset of data, while not exhausting available disk resources. A full staging and processing of certain streams could then be decided, when analyses have reached a sufficient stage of preparation. Further careful thought would be needed regarding how to distribute and manage resources in a more tape-reliant model so as not to introduce bottlenecks.

Another challenge that will be magnified by the Upgrade~II data volumes is the overhead of reading the sprucing output from disk in user jobs when creating tuples. Storage formats should be designed with access patterns and hardware constraints in mind to ensure that analyses can make efficient use of resources.

To maximally benefit the analysis process, tuples should be easy to manipulate for the purpose which they are intended for and not burdensome. This generally means ensuring they are not too large, and contain the right amount of observables for their purpose. The currently prevailing workflow is for analysts to process a given stream to produce their tuples on demand, using WLCG compute and disk resources. Any additional processing passes are carried out by the user on typically interactive nodes after transferring the output from WLCG storage.

With much higher data rates and volumes available offline in Upgrade~II, tuples will become much larger and more difficult to manipulate and especially transfer over the network in reasonable time. Keeping the overhead of reading spruced streams low means minimising the passes over these streams, but this results in larger, more unwieldy tuples. This potentially necessitates following a different approach:
\begin{enumerate}
    \item[(1)] Minimise the passes over a given stream, by creating as many tuples as needed according to demand in a single pass (may involve one or more related or unrelated analyses).
    \item[(2)] Run additional processing passes on the resulting tuples to ``mold" the data according to the analysis purpose (may involve actual analysis steps).
\end{enumerate}
Step (1) could be organised and prepared formally or informally. Tuples will be large, difficult to analyse, and in their most general form covering a wide range of analysis goals and expected studies.
Step (2) involves passes that transform these large tuples into a form that is easier to manipulate for user analysis. These could involve steps such as splitting the tuples out into multiple, smaller files to make them quicker to read, filtering tuples to reduce their size ({\it i.e.}\ carrying out a selection), or transforming tuples into {\it e.g.}\ a smaller size tuple with less observables, or some other objects, like histograms.
This would effectively encourage heavy processing and analysis steps to be performed by analysts using WLCG resources. This is made possible using the Analysis Productions framework~\cite{Skidmore:2827261} being used in Run~3 tuple production. The analysis of the full Run~3 data will be an ideal benchmark to test the scalability of this solution. This may constitute the beginning of an ``analysis facility", where users may scale up their analysis workflows by leveraging WLCG resources. It is expected that additional functionalities and optimisations will be needed for the Upgrade~II data volumes. Other hardware backends might also need to be integrated in such workflow to enable processing of the huge Upgrade~II samples.

The options available for additional processing passes can be augmented with pre-defined workflows or workflow templates such as: adding resampled particle-identification variables or reweighting to the simulation tuples, among many other commonly performed analysis tasks within \lhcb, which may save time while avoiding individual user errors and the duplication of effort.


The data sprucing, tuple production and data analysis workflows are all executed on the resources provided by the WLCG sites. \lhcb interacts with the provided compute and storage resources through the Workload Management and Data Management components of the LHCbDIRAC services~\cite{DiracCHEP2020}. These services will require enhancements to support the increase in data rates and CPU needs expected in Upgrade~II. In addition, these services will need to support the evolution of the WLCG services themselves (compute, disk and tape storage, authentication, security, {\it etc.}), as well as modern architectures as the need arises. 

The computing environments are expected to continue to evolve rapidly, and the tooling and workflows will need to be adapted to these changes. The resources and services provided by WLCG will continue to be used in the Upgrade~II phase, but R\&D effort will be needed to optimize the use of these resources. This does not only concern the traditional resources such as CPU time and storage sizes, but also quantities such as memory utilization, network bandwidth, latency, and storage bandwidth that affect throughput and performance of the processing jobs.  This broad range of metrics needs to be considered to ensure timely execution of the workflows. In addition, the evolution of user analysis environments and tools will require R\&D efforts for ``analysis facilities" infrastructures, to offer required functionality and performance.

The need to improve the sustainability of the HEP computing infrastructures, and lower its carbon footprint also has implications for the LHCb jobs run on the WLCG. In particular, the power consumed by the various workflows should be studied and improved. This requires a sustained effort of all the software developers in the collaboration, as well as a close collaboration with WLCG and its sites.

\section{Closing remarks}
\label{sec:closing}

The LHCb Upgrade~II programme offers a unique opportunity to fully exploit the physics capabilities of the High-Luminosity LHC. This poses however an unprecedented software and computing challenge, in order to enable the readout, reconstruction, selection and analysis of the largest ever data bandwidth in a HEP experiment.
The first upgrade of the LHCb experiment has set the ground to overcome such a challenge, but an intense R\&D programme is still imperative to fulfil the requirements of the demanding Upgrade~II conditions. This will largely benefit the community beyond LHCb as future experiments will face similar hurdles.
To ensure a strong R\&D programme in software and computing over the next decade, two ingredients are essential: software training and career paths for highly skilled researchers and engineers in this area.

The software of current experiments has largely evolved during recent years, with experiment-specific challenges requiring more technical and customised solutions. Important efforts have also been dedicated to develop common solutions in areas with common challenges, such as storage. Both approaches require deep knowledge of computing and programming. The next generation of experiments will require even more ad-hoc solutions to manage the unprecedented data rates. This will require close collaboration between physicists, engineers and software developers. The software skills learnt in the average physics (master) degrees across Europe are largely insufficient to design and develop such software, and consequently specific training is required for researchers working in this domain. There has been progress on this in recent years, with initiatives such as those led by the HEP Software Foundation~\cite{stewart_2025_15097160} giving a base for common needs. However, with the specialisation of the software, more specific trainings are required for communities working with particular solutions. LHCb plans a continuous effort in training as part of the Upgrade~II R\&D given the fast evolution of computing technologies but would also benefit from cross-experiment initiatives where appropriate. 

The specialisation and higher complexity of the software solutions also require a full and continued dedication of researchers with the appropriate skills to their design, development and maintenance. A model based on the dedication of a small fraction of researcher time to software developments and with a large turnover of people is not sustainable any more, given the highly technical skills needed and growing complexity of the tasks. It is therefore imperative to ensure the stabilisation of researchers that have acquired the necessary knowledge and expertise in the field. Clear career paths should be defined both in research centres and universities. Moreover, it is of utmost importance to offer the right balance between the necessary tasks related to maintenance of existing software frameworks and the freedom to explore future solutions and R\&D programmes, given the fast evolution of the computing technology in recent and future years.

These points require a strong investment of European institutions on software training and software- and computing-oriented career paths within HEP research groups. This will enable not only the necessary R\&D for the challenges of the next decade, but will also place European groups in a leading position to address the challenges that will arise further on with the next generation of colliders and experiments. The R\&D on software and computing for LHCb Upgrade~II will play a major role in this endeavour.


\addcontentsline{toc}{section}{References}
\bibliographystyle{LHCb}
\bibliography{main,standard,LHCb-PAPER,LHCb-CONF,LHCb-DP,LHCb-TDR}
 


\end{document}